\documentclass[showpacs,aps,prb,twocolumn,10pt,longbibliography]{revtex4-2}
\usepackage{graphicx}

\usepackage{bm}
\usepackage{color}
\usepackage[colorlinks=true, allcolors=blue]{hyperref}
\usepackage{amsmath}
\usepackage{amssymb}
\usepackage{enumerate}
\usepackage{xspace}
\usepackage{mathrsfs}
\usepackage{mathptmx}
\usepackage[utf8]{inputenc}
\usepackage[dvipsnames]{xcolor}

\begin{document}

\title{Bridging atomistic spin dynamics methods and phenomenological models of single pulse ultrafast switching in ferrimagnets}

\author{Florian~Jakobs}
\affiliation{Dahlem Center for Complex Quantum Systems and Fachbereich Physik,  Freie Universit\"{a}t Berlin,  14195 Berlin, Germany}

\author{Unai~Atxitia}
\affiliation{Dahlem Center for Complex Quantum Systems and Fachbereich Physik,  Freie Universit\"{a}t Berlin,  14195 Berlin, Germany}

\begin{abstract}
 We bridge an essential knowledge gap on the understanding of all-optical ultrafast switching in ferrimagnets; namely, the connection between atomistic spin dynamics methods and macroscopic phenomenological models.   
All-optical switching of the magnetization occurs after the application of a single femtosecond laser pulse to specific ferrimagnetic compounds.  This strong excitation puts the involved degrees of freedom, electrons, lattice and spins out-of-equilibrium between each other. 
Atomistic spin models have quantitatively described all-optical switching in a wide range of experimental conditions, while having failed to provide a simple picture of the switching process. Phenomenological models are able to qualitatively describe the dynamics of the switching process. However, a unified theoretical framework is missing that describes the element-specific spin dynamics as atomistic spin models with the simplicity of phenomenology. Here, we bridge this gap and present an element-specific macrospin dynamical model which fully agrees with atomistic spin dynamics simulations and symmetry considerations of the phenomenological models. 
\end{abstract}
\maketitle

\section{Introduction}

Since its experimental discovery \cite{Stanciu2007}, the theoretical description of laser induced all-optical switching (AOS) of the magnetization in GdFeCo  ferrimagnetic alloys has remained a challenge.
Despite intense experimental and theoretical research in the field \cite{Stanciu2007,Vahaplar2009,Radu2011,Ostler2012,Mentink2012,LeGuyader2012,Graves2013,Barker2013,Baryakhtar2013,Gridnev2016,Schellekens2013,Mangin2014}, an established and unified picture of the process is still missing. 
Experimental findings are mostly compared or interpreted in terms of atomistic spin dynamics simulations \cite{Ostler2011,Wienholdt2013,Chimata2015,Jakobs2021,Ceballos2021}, multisublattice spin dynamics based on symmetry arguments \cite{Mentink2012,Radu2015,Mentink2017}, and based on the Landau-Lifshitz-Bloch equation \cite{Atxitia2012,Atxitia2013,Atxitia2014}.
The main goal of the present work is the revision, extension and merging of these approaches into a unified model.

Atomistic spin dynamics (ASD) models have been used before to quantitatively describe ultrafast dynamics in $3d$ transition metals \cite{Zahn2021,Zahn2022} and $4f$ rare-earth ferromagnets \cite{Frietsch2015,Frietsch2020}.  
They have also been used in GdFeCo, to describe the equilibrium thermal properties~\cite{Ostler2011}, the thermal character of AOS \cite{Ostler2012}, the so-called transient ferromagnetic-like state \cite{Radu2011}, the demonstration of spin-current-mediated rapid magnon localisation and coalescence \cite{IacoccaNatComm2019} and the possibility of AOS using picosecond-long laser pulses \cite{Jakobs2021}. 
Results from atomistic spin models also compare qualitatively well to an analytical theory based on the excitation of spin-wave exchange modes~\cite{Barker2013}, provide insights for optimal electron, phonon and magnetic characteristics for low energy switching~\cite{Atxitia2015} and predict maximum repetition rate using two consecutive laser pulses \cite{Atxitia2018}. More sophisticated, orbital-resolved atomistic models  provide insights on the role of the intra-exchange coupling between $4f$ and $5d$ electrons in the dynamics of GdFeCo alloys\cite{Wienholdt2013}. Atomistic models can naturally describe switching in Gd/Fe multilayers composed of very thin layers \cite{Xu2016,Gerlach2017}. 
Recent observations \cite{Banerjee2020,Banerjee2021} of single pulse switching in Mn$_2$Ru$_x$Ga alloys are also well-described by ASD methods \cite{Jakobs2022}.
Despite the demonstrated success in modeling AOS, ASD simulation results are cumbersome to interpret without an analytical model that unveils the role of the different processes and interactions during the switching process. This potential semi-analytical model has to capture most of the features of the ASD simulations.

Semi-phenomenological models describing switching already exist. 
A macroscopic theory for the description of the dynamics and relaxation of the macroscopic (sublattice) magnetization of ferromagnets and antiferromagnets was developed originally by Baryakhtar \cite{BaryakhtarReview,Baryakhtar2013}. An extension of such phenomenology to ferrimagnets in the context of ultrafast spin dynamics was introduced in Ref. \cite{Mentink2012}.
At the ultrafast scale, magnetization dynamics are dominated by atomic scale spin excitations, these spin dynamics are driven by dissipative processes which in ferrimagnets are two-fold, relativistic and exchange driven. Relativistic processes allow for exchange of angular momentum between the spins and lattice degree of freedom due to the presence of spin-orbit interaction connecting them. Exchange processes can arise due to transport of spin angular momentum -- spin and magnon transport --  which is the only mean to exchange angular momentum in ferromagnets. In multisublattice magnets another, different pathway opens, namely, local exchange of angular momentum.
To account for such local exchange processes in  ferrimagnets, the equation of motion for the magnetization dynamics proposed by Landau and Lifshitz \cite{Landau1935} is enhanced by an exchange relaxation term \cite{Mentink2012,Mentink2017,Baryakhtar2013,Kamra2018}. 
Within this macroscopic model, the exchange relaxation dominates the dynamics when the magnetic sublattices are driven into mutual non-equilibrium.  
Qualitative agreement to experiments in two-sublattice magnets has been demonstrated \cite{Mentink2017}, such as AOS in ferrimagnetic GdFeCo using fs laser pulses \cite{Mentink2012} and ps laser pulses \cite{Davies2020}, AOS in Heusler semimetals Mn$_2$Ru$_x$Ga \cite{Davies2020b}, or element-specific demagnetization of ferromagnetic NiFe alloys \cite{Radu2015}. 
Quantitative comparison of this model to neither experiments nor ASD simulations have been conducted so far. 
While the arguments behind such phenomenology are robust, the range of applicability and the validity of the model parameters could be questioned. 
For instance, the parameters defining the relativistic and exchange relaxation are assumed to be constant and of the same order. The magnetic free energy functional is calculated for near thermal equilibrium states. This implies a relatively strong coupling to the heat-bath, while switching conditions are supposedly fulfilled when exchange relaxation between sublattices dominates over the relaxation to the heat-bath. 

An alternative macroscopic model directly derived from an atomistic spin model has also been proposed. 
This model is  based in the Landau-Lifshitz-Bloch (LLB) equation of motion \cite{Garanin1997,Chubykalo-Fesenko2006,Atxitia2012,Nieves2014,Atxitia2016}. 
The LLB model for two-sublattice magnets \cite{Atxitia2012,Nieves2014} has been used in the context of AOS in GdFeCo, e.g. the element-specific demagnetization rates compare well to experiment, and it predicts that near the magnetic phase transition the otherwise slower Gd sublattice becomes faster than Fe \cite{Atxitia2014}, as recently observed \cite{Hennecke2019}. 
The LLB model has been demonstrated to provide accurate analytical expressions for the temperature dependence of the relativistic relaxation parameter as well as for the non-equilibrium effective fields below and above the critical temperature \cite{Nieves2014}. 
 Moreover, the LLB model also describes the transverse motion of the magnetization. This makes it the preferred model for computer simulations of  heat-assisted magnetic recording \cite{Vogler2019} and realistic description of all-optical switching \cite{RaposoPRB2022}, and ultrafast spintronics, such as domain wall motion \cite{Schlickeiser2014,Moretti2017} or skyrmion creation by ultrafast laser pulses \cite{Lepadatu2020}. So far the LLB model and Baryakhtar-like models have been considered as complementary approaches. Here, we merge them into one unified approach.

In this work we address  the issues discussed above by directly comparing both phenomenological models to ASD simulations. We do so since ASD simulations have been already quantitatively compared to experiments in literature. We find that quantitative comparison between ASD and both phenomenological models is partially possible for laser excitation producing small deviation from equilibrium. 
However, those models hardly reproduce magnetic switching using the same parameter values describing the relaxation of small perturbations.  
Here, based upon those phenomenological models, we propose a macroscopic model that compares precisely to the magnetization dynamics calculated using ASD simulations, including element-specific magnetization relaxation and switching. 
This model bridges atomistic spin dynamics based models and previously proposed phenomenological models. Notably, it provides a deeper understanding to the parameters entering the phenomenological models and sheds some light into the process of ultrafast switching in ferrimagnets.

The work is broken down in the following way: in Sec. \ref{sec:secII}, we present the atomistic spin model for the calculation of the magnetic equilibrium properties and non-equilibrium dynamics. The equilibrium properties are compared to a mean field model. We then provide atomistic calculations of the ultrafast magnetization dynamics with input from the two temperature model. These results are the basis for the comparison to the phenomenological models presented in Sec. \ref{sec:secIII}. Firstly, we present the Baryakhtar model and the Landau-Lifshitz-Bloch model. Secondly, we compare the ultrafast magnetization dynamics calculated with those models to the atomistic spin dynamics results. Finally, in Sec. \ref{sec:secIV} we present the unified phenomenological model, a hybrid model combining Baryakhtar and LLB models, and its comparison to atomistic spin dynamics.

\section{Atomistic Spin Model}
\label{sec:secII}
 Ferrimagnetic materials characterise by spontaneous magnetization as a resultant of two or more components of non-parallel magnetic moments \cite{Barker2021}.  
Atomistic spin models based on the Heisenberg Hamiltonian can be considered one of the simplest microscopic models able to reproduce the equilibrium properties of ferrimagnets. 
The spin system energy due to only the exchange interactions can be described by an effective Heisenberg model:
 \begin{eqnarray}  
\mathcal{H}= - \sum_{i \neq j} J_{a} \mathbf{S}_{a,i} \cdot \mathbf{S}_{a,j}
- \sum_{i \neq j} J_{b} \mathbf{S}_{b,i} \cdot \mathbf{S}_{b,j}
- \sum_{i \neq j} J_{ab} \mathbf{S}_{a,i} \cdot \mathbf{S}_{b,j}
\label{eq:Ham}
\end{eqnarray}
where $J_{a(b)(ab)}$ is the exchange constant between neighboring sites represented by two classical spin vectors 
$\mathbf{S}_i$ and $\mathbf{S}_j$ ($|\mathbf{S}|=1$). Further, one can include magnetic anisotropy terms to Eq. \eqref{eq:Ham} to set a preferential axis for the magnetization. 
However, since the anisotropy energy is relatively low it  plays a marginal role in the switching process. This makes for a simpler Hamiltonian and a more direct comparison to the phenomenological models.
To model a ferrimagnet, one needs to consider two alternating sublattices of unequal and antiparallel moments, with three exchange coupling constants: 
ferromagnetic for each sublattice ($J_a$ and $J_b$) and a third for the antiferromagnetic interaction between them, $J_{ab}$.
For instance, GdFeCo alloys are composed of a transition metal FeCo and a Gd rare-earth sublattices. 
We model the Fe and Co  spins as only one magnetic sublattice, and we assume a common  atomic magnetic moment of $\mu_{\rm{FeCo}}=1.94\mu_{\textrm{B}}$.  
In these alloys the rare-earth impurities add localised $4f$ spins to the system assumed to be, $\mu_{\rm{Gd}}=7.6\mu_{\textrm{B}}$. 
The amorphous nature of GdFeCo is modelled by using a simple cubic lattice model but with random placements of Gd moments within the lattice to the desired concentration.
The applicability of the Heisenberg approximation relies on the stability of local moments under rotation and at high temperature where Stoner excitations are generally weak \cite{Chimata2012}. It is assumed that the electronic properties are temperature-independent in the range where the system is magnetically ordered.

\subsection{Atomistic spin dynamics}

Equilibrium and non-equilibrium element specific magnetic properties of a ferrimagnet are calculated using atomistic spin dynamics simulations which are based in the  stochastic-Landau-Lifshitz-Gilbert equation (s-LLG) \cite{Nowak2007BOOK}
\begin{eqnarray}  
           (1+\lambda_i^2)\mu_{s,i} \dot{\mathbf{S}}_i =
              - \gamma  \mathbf{S}_i  \times  \left[ \mathbf{H}_i 
             -  \lambda_i \;   \left( \mathbf{S}_i  \times \mathbf{H}_i  \right) \right],
             \label{eq:sLLG}
\end{eqnarray}
where $\gamma$ is the gyromagnetic ratio, and $\lambda_i$ is the so-called phenomenological sublattice specific damping parameter. 
By including a Langevin thermostat the spin dynamics including statistical -- equilibrium and non-equilibrium thermodynamic properties can be obtained.  An effective field-like stochastic term $ \boldsymbol{\zeta}_i$ is added to the effective field $\mathbf{H}_i= \boldsymbol{\zeta}_i(t) - \frac{\partial \mathcal{H}}{\partial \mathbf{S}_i}$, with   white noise properties~\cite{Atxitia2009}: $
\langle \boldsymbol{\zeta}_i(t) \rangle = 0 \quad \text{and} \quad \langle \boldsymbol{\zeta}_i(0) \boldsymbol{\zeta}_j(t) \rangle = 2 \lambda_i k_\text{B} T \mu_{s,i} \delta_{ij}\delta(t)/\gamma.$ 
The variance of the Langevin noise is chosen such that the fluctuation-dissipation theorem is full filled.

 \begin{figure}[!tb]
\includegraphics[width=8.5cm]{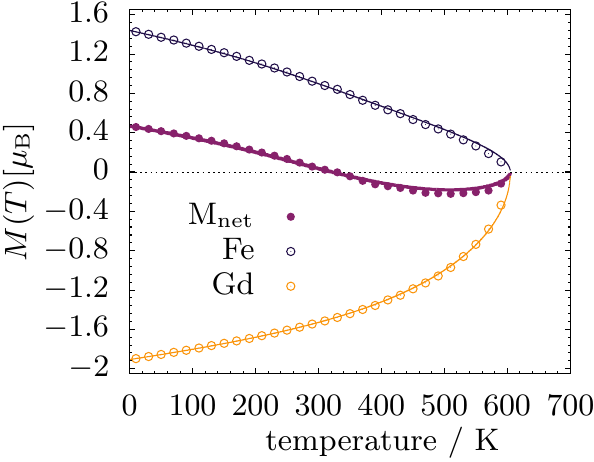}
\caption{Equilibrium magnetization of a GdFeCo alloy for Gd concentration, $x_{\rm{Gd}}=25\%$. Element-specific normalized equilibrium magnetization and net equilibrium magnetization, $M(T)=x_{\rm{Gd}} \mu_{\rm{Gd}} m_{\rm{Gd}}-x_{\rm{Fe}} \mu_{\rm{Fe}} m_{\rm{Fe}}$, where $\mu_{\rm{Gd(Fe)}}$ is the atomic magnetic moment of Gd(Fe).  Lines correspond to the mean-field approximation with renormalized exchange parameters. Symbols correspond to atomistic spin dynamics simulations.}
\label{fig:ASD-MFA-M}
\end{figure}

\subsection{Mean-field approximation}
Exact analytical expressions for the $M(T)$ curve are cumbersome to derive due to the many body character of the problem. 
Here we resort the mean field approximation (MFA), already used in previous works \cite{Ostler2011,Barker2013,Hinzke2015}. 
We note that to be able to apply the MFA for the GdFeCo impurity model, and thus translation non-symmetric with respect to  spin variables  $\mathbf{S}_i$, we need to transform the Heisenberg Hamiltonian to a symmetric one. 
We use the spin analogy of the virtual crystal approximation (VCA) to transform the disordered lattice Hamiltonian $\mathcal{H}$ to a symmetric VCA Hamiltonian $\mathcal{H}_{\mathrm{VCA}}$.
Within the VCA we evaluate the effective sublattice exchange parameters, given by the sum of the exchange interactions of a given spin at a site $\mathbf{r}_i$ of sublattice $i$ with all other atoms of this sublattice.
This involves weighting the exchange parameters by the relative composition, $x_i \equiv \mathrm{concentration \ species} \ i$ \cite{Barker2013},  
\begin{equation}
J_i= \sum_{\mathbf{r}_i,\mathbf{r}'_i} J(\mathbf{r}_i,\mathbf{r}'_i) \underbrace{\equiv}_{\mathrm{VCA}} 
\ \  x_i  J(\mathbf{r}_i,\mathbf{r}'_i)    \  \  \   \mathrm{intrasublattice}
\end{equation}
whereas the intersublattice effective exchange reads 
\begin{equation}
J_{ij}= \sum_{\mathbf{r}_i,\mathbf{r}'_j \notin A_i} J(\mathbf{r}_i,\mathbf{r}'_j) \underbrace{\equiv}_{\mathrm{VCA}} 
\ \  x_i  J(\mathbf{r}_i,\mathbf{r}'_j)    \  \  \   \mathrm{intersublattice}
\end{equation}
Thus the VCA Hamiltonian  reads 
 \begin{equation}
 \label{eq:VCAHamiltonianSI}
 \mathcal{H}_{\mathrm{VCA}}= \sum_{j \in A_i} J_i \mathbf{S}_i \cdot \mathbf{S}_j+
                                                  \sum_{j \notin A_i} J_{ij} \mathbf{S}_i \cdot \mathbf{S}_j
 \end{equation}
 where $A_i$ represent the magnetic sublattice of the spin $\mathbf{S}_i$.   
In the exchange approximation we define the MFA field as
\begin{equation}
\mu_a H_a^{\rm{MFA}} = z_a J_{aa} m_a + z_{ab} J_{ab} m_b 
\label{eq:MFA-LLB}
\end{equation}
The element-specific equilibrium magnetization is calculated via the self-consistent solution of $m_a=L(\beta \mu_a H^{\rm{MFA}}_a)$ and  $m_b=L(\beta \mu_b H^{\rm{MFA}}_b)$. 
$z_a$ and  $z_{ab}$ correspond to the number of first nearest neighbours of type $a$ and $b$, respectively.
It is well-known that the MFA overestimates the value of the critical temperature $T_C$. 
However, a very good agreement between ASD and MFA can be obtained by using a reduced value for the exchange parameters, even for multilattice magnets \cite{Hinzke2015}.
Figure \ref{fig:ASD-MFA-M} shows element-specific $M_a=x_{a}\mu_{a} m_{a}(T)$ using ASD simulations and renornalized MFA for $x_{\rm{Gd}}=25\%$. 
Net magnetization is also shown in Fig. \ref{fig:ASD-MFA-M}, which is defined as $M(T)=x_{\rm{Gd}}\mu_{\rm{Gd}} m_{\rm{Gd}} - x_{\rm{Fe}}\mu_{\rm{Fe}} m_{\rm{Fe}}$.
The agreement between ASD and MFA is good enough for all the temperature regions. 
We observe the presence of compensation temperature $T_{\rm{M}}$ at room temperature for 
$x_{\rm{Gd}}=25 \%$ at which the thermally average magnetization of both sublattices are equal but opposite, so that the magnetization of the system is equal to zero $M(T_{\rm{M}}) = 0$. The mapping of the atomistic spin model and the corresponding mean-field approximation turns out to be necessary for a quantitative comparison to the phenomenological models, and thereby paramount for the unification of both pictures. 

\subsection{Two Temperature Model}

 Single pulse all-optical switching has been demonstrated to be a thermal process in ferrimagnetic GdFeCo alloys \cite{Ostler2012} and in Mn$_2$Ru$_x$Ga Heusler semi-metals \cite{Banerjee2020}. 
 Ultrafast heating by optical or electric means are sufficient to achieve switching in specific GdFeCo alloys \cite{Yang2017}. 
 Although the minimum achievable duration of the electric pulses are limited to picoseconds, those are better suited for potential integration  into applications. 
 Laser pulses can be as short as only a few femtoseconds, which permits to excite the electron system in timescales of the order of the exchange interaction allowing for the investigation of fundamental physics governing switching. In this work, we center in excitation of the ferrimagnetic GdFeCo using femtosecond laser pulses.      
 When a metallic ferrimagnetic thin film is subjected to a near infrared laser pulse, only the electrons are accessible by the photon electric field. Initially, the absorbed energy is barely transferred to the lattice and consequently the electron system heats up. 
 The electron and phonon temperatures are decoupled for up to several picoseconds until the electron-phonon interaction equilibrates the two heat-baths. 
 This phenomenology is well captured by the so-called two-temperature model (2TM) \cite{Kaganov1957,Chen2006}  which can be written as two coupled differential equations:
\begin{align}
C_{\rm{el}} \frac{\partial T_{\rm{el}}}{\partial t} &= -g_{\rm{ep}}\left( T_{\rm{el}} - T_{\rm{ph}} \right) + P_{l}(t)
\label{eq:2TM-el}
\\
C_{\rm{ph}} \frac{\partial T_{\rm{ph}}}{\partial t} &= +g_{\rm{ep}}\left( T_{\rm{el}} - T_{\rm{ph}} \right).
\label{eq:2TM-ph}
\end{align}
$C_{\rm{el}}=\gamma_{\rm{el}} T_{\rm{el}}$ where  $\gamma_{\rm{el}}=6\times 10^2$ J/m$^3$K$^2$, and $C_\text{ph}=3.8\times 10^6$ J/m$^3$K represent the specific heat of the electron- and phonon system.
The electron-phonon coupling is taken temperature independent, $G_{\rm{ep}}=7 \times 10^{17}$ J/m$^3$K.
Here, $P(t)$ is a Gaussian shaped pulse with a duration of 55 fs.
The exact values of the parameters entering the TTM in GdFeCo are still unknown. The values we use here are close to the commonly used, e.g. Refs. \cite{Barker2013,Ostler2012,Jakobs2022}.  


\subsection{Ultrafast magnetization dynamics using ASD}

\begin{figure}[!tb]
\includegraphics[width=8.25cm]{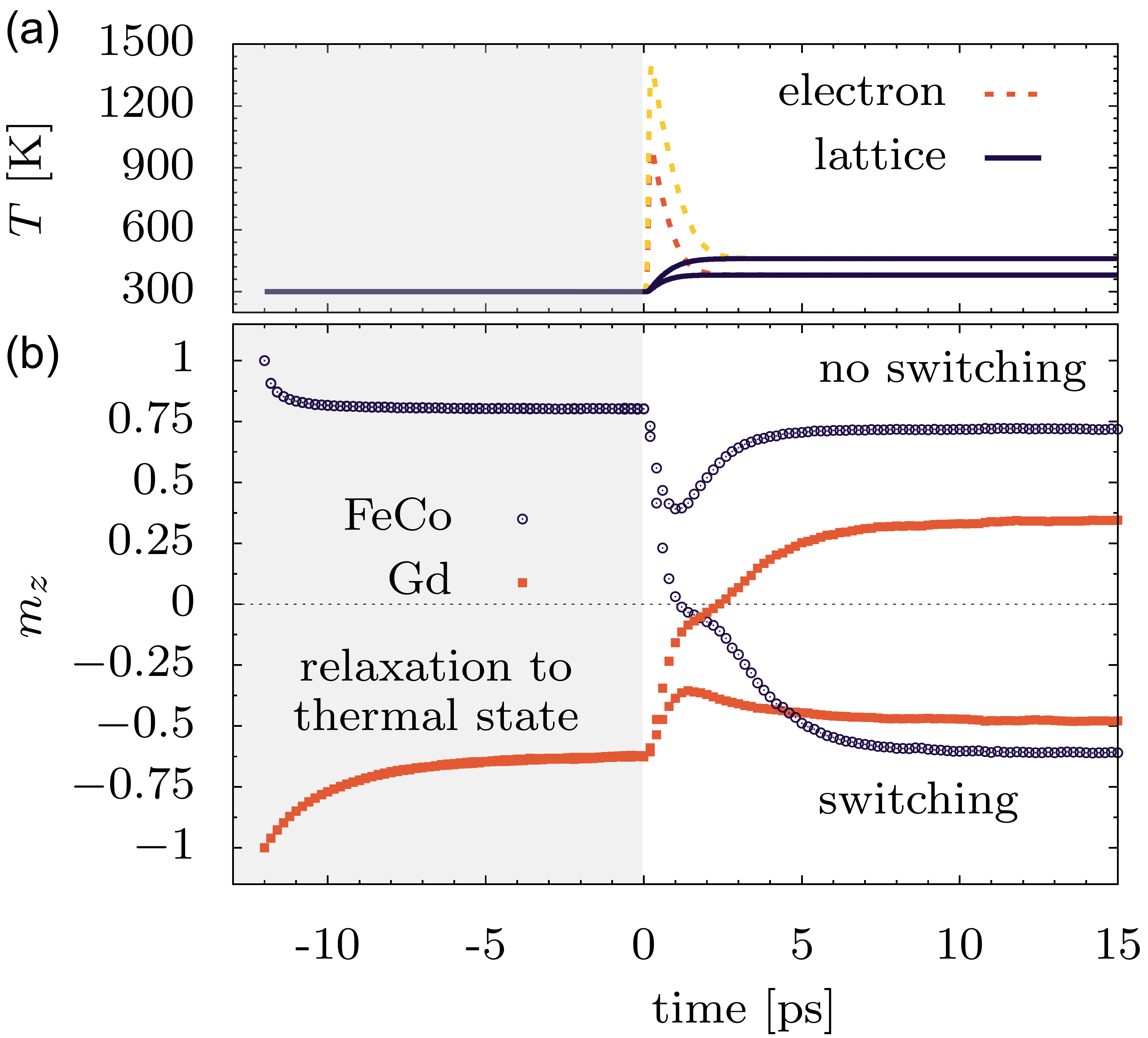}
\caption{ (a) Electron and lattice temperature dynamics for two laser pulse power values, $P_0$ and $2P_0$. 
Both electron and lattice temperature are kept constant, $T=300$K, for  $t<0$. At $t=0$ a laser pulse is applied and the dynamics of the electron and lattice temperature heat up. The dynamics of those temperatures are theoretically described by the two-temperature model. 
(b) Element-specific magnetization dynamics induced by the heat profile at (a). The dynamics are calculated using atomistic spin dynamics methods. For lower laser powers $P_0$, the magnetization of both sublattices demagnetize rapidly and remagnetize towards the new equilibrium. For laser power $2P_0$, the magnetization of both sublattices demagnetizes and switches. After switching they relax towards the thermal equilibrium state. 
GdFeCo alloys with $x_{\rm{Gd}}=25 \%$ are calculated.}
\label{fig:ASD-switching}
\end{figure}

Element-specific magnetization dynamics induced by a femtosecond laser pulse are calculated by combining the atomistic s-LLG equation for the spin dynamics (Eq. \eqref{eq:sLLG}) and the 2TM for the electron temperature (Eq. \eqref{eq:2TM-el}). The electron system acts as heat-bath for the atomic spins.
We consider a lattice with $N=50\times 50\times 50$ spins, and  
damping parameters, $\lambda_{\rm{Gd}}=0.01=\lambda_{\rm{Fe}}$. 
Figure \eqref{fig:ASD-switching} shows, for $t<0$, the dynamics of the element-specific magnetization from an initial saturated state ($T=0$ K), towards thermal equilibrium with the heat-bath which is set to $T=300$ K. The relaxation dynamics of Fe sublattice is faster than those of the Gd sublattice. 
This comes out naturally as the element-specific dissipation of angular momentum scales as $\dot{m}_z \sim \gamma \lambda/\mu_{s}$, in Gd ($\mu_{\rm{Gd}}=7.6 \mu_{\rm{B}}$) is slower than in Fe sublattice (\textbf{$\mu_{\rm{Gd}}=1.94 \mu_{\rm{B}}$}). 
Once the magnetic system is in thermal equilibrium with the heat-bath, we apply the laser pulse, $t>0$, which introduces energy into the electron system and induces ultrafast magnetization dynamics.
To illustrate the switching and no switching dynamics we consider two limiting cases, dynamics induced by low laser power, $P_0$, and large laser power, $2P_0$.
The electron temperature increases up and above the Curie temperature in time scales of a few hundreds of femtoseconds
Fig. \eqref{fig:ASD-switching} (a). 
This reflects in the magnetic system as a fast demagnetization of both Fe and Gd sublattices.
For relatively low laser power, $P_0$, the magnetization of both sublattices reduces while the electron temperature remains relatively high. 
Once the electron temperature reduces and equalizes to the lattice temperature, the magnetization recovers to the thermal state given by the heat-bath temperature, which is higher than initially ($T=300$ K). 
This is why the final magnetization value is smaller than the initial one. 
For higher laser powers, $2P_0$, the magnetization of both sublattices reduces quickly. 
The Fe sublattice faster than the Gd one. 
Once the magnetization of the Fe sublattice hits zero, instead of remaining demagnetized, the magnetization starts to develop toward the opposite direction, while the magnetization of the Gd sublattice is still in the process of demagnetization. 
During a couple of picoseconds, both sublattice magnetization are aligned along the same direction, similar to a ferromagnet. Consequently, this non-equilibrium state has been named the transient ferromagnetic-like state \cite{Radu2011}. 
One can observe in Fig. \eqref{fig:ASD-switching} (b) that the demagnetization rates of both sublattices slow down when the  Fe magnetization crosses zero.
This change reveals the set in of a process driving the 
magnetization dynamics different to the one driving the initial demagnetization. 
It has been argued that at this point direct exchange of angular momentum between sublattices dominates over processes of relativistic origin, which in turn dissipate angular momentum into the heat-bath. 
Interestingly, soon after switching, both sublattice magnetization rapidly relax to equilibrium indicating that relaxation into the heat-bath dominates the dynamics.

\section{Phenomenological models}
\label{sec:secIII}

Differently to ASD simulations, phenomenological models describe the element-specific magnetization dynamics by solving  two coupled equations of motion, one for each sublattice. In this work we aim at finding a phenomenological model that describes the same element-specific magnetization dynamics as those coming out from the ASD simulations (Fig. \ref{fig:ASD-switching}). 
The starting point is the comparison of the ASD simulations to well-known phenomenological models. We show that those models are unable to describe in a satisfactory way the different  element-specific magnetization dynamics studied in the previous section and summarized in Fig. \ref{fig:ASD-switching}. 

\subsection{Baryakhtar model}

The simplest model to describe element-specific magnetization dynamics and switching in ferrimagnets was proposed by Mentink and co-workers \cite{Mentink2012}. Longitudinal spin dynamics was derived from Onsager’s relations
 \begin{eqnarray}  
\frac{\mu_{a}}{\gamma_a}\frac{dm_a}{dt} & = &  \alpha^{\rm{B}}_a \mu_{a}H_a + \alpha^{\rm{B}}_e (\mu_{a}H_a - \mu_{b}H_b)
\label{eq:Onsager-multisublattice-a}\\
\frac{\mu_{b}}{\gamma_b}\frac{dm_b}{dt} & = &  \alpha^{\rm{B}}_b \mu_{b}H_b + \alpha^{\rm{B}}_e (\mu_{b}H_b - \mu_{a}H_a) 
\label{eq:Onsager-multisublattice-b}
\end{eqnarray}
here, 
$\alpha^{\rm{B}}_{a,b}$ stands for the relaxation parameter of relativistic origin, which dissipates angular momentum out of the spin system, and $\alpha^{\rm{B}}_e$ stands for the exchange relaxation parameter and describes the rate of dissipation of angular momentum between sublattices. By construction exchange relaxation conserves the total angular momentum. 
We emphasize here the difference in the notation between the atomic relaxation parameter, $\lambda$, describing the dissipation of the atomic spins in ASD simulations and the macrospin relaxation parameter, $\alpha$, describing the dissipation of the whole magnetic sample.
Within this model, the values for $\alpha^{\rm{B}}_{a,b}$  and $\alpha^{\rm{B}}_e$ are  unknown but used as fitting parameters when compared to experiments.
The internal effective field $H_{a(b)}$, acting on sublattice $a(b)$ are derived from a  non-equilibrium  mean-field approximation, 
 \begin{eqnarray}  
\mu_a H_{a}= -\beta^{-1} L^{-1} (m_{a}) + \mu_a H_a^{\rm{MFA}} 
\label{eq:effective-fields-MFA-Mentink}
\end{eqnarray}
where, $L^{-1}(x)$ is the inverse Langevin function, $\beta=1/k_B T$, where $T$ represents the temperature of the heat-bath to which the spin system is coupled to. At equilibrium, the effective field is $H_a=0$, as $m_a=L(\beta\mu_a H_a^{\rm{MFA}})$. The same arguments apply for sublattice $b$. 
It turns out that by solving Eqs. \eqref{eq:Onsager-multisublattice-a} and \eqref{eq:Onsager-multisublattice-b} together with the 2TM, described in Eqs. \eqref{eq:2TM-el} and \eqref{eq:2TM-ph}, one obtains similar ultrafast magnetization dynamics as those using ASD simulations (Fig. \eqref{fig:ASD-switching}). 
Element-specific demagnetization \cite{Radu2015}  and switching dynamics \cite{Mentink2017} based on this approach have been discussed thoughtfully before. 
On those works, the values for the relaxation parameters, relativistic and exchange, are taken constant and of the same order, $\alpha^{\rm{B}}_{\rm{Fe}}\approx \alpha^{\rm{B}}_{\rm{Gd}}\approx \alpha^{\rm{B}}_{e}$. 
We note that here $\alpha^{\rm{B}}_a$ defines the rate of change of angular momentum ($m\mu/ \gamma$). 
It differs from the definition of intrinsic damping parameters in ASD, which are related to the rate of change of the magnetization ($m$). 
Similarly to ASD methods though, within the Baryakhtar model the observed fast dynamics of the Fe sublattice is related to a smaller value of atomic magnetic moment.  

The switching process within the Baryakhtar-like model is explained in the following manner. 
Since the Fe sublattice reacts faster than Gd to heating it is expected to remain closer to thermal equilibrium with the heat-bath. 
This translates into a smaller non-equilibrium effective field acting on Fe than in Gd,  $H_{\rm{Fe}}\ll H_{\rm{Gd}}$, during the action of the laser pulse. 
For strong enough pulses, the Fe magnetization rapidly reduces, $m_{\rm{Fe}}\approx 0$, still $H_{\rm{Fe}}$ is small in comparison to $H_{\rm{Gd}}$, in a way that the dynamics of Fe can be fairly approximated by $\dot{m}_{\rm{Fe}} \approx \alpha^{\rm{B}}_{e} H_{\rm{Gd}}$. 
This drives the magnetization of Fe towards the opposite direction.
The field, $H_{\rm{Gd}}$ is defined by the energy of the system, $H^{\rm{MFA}}_{\rm{Gd}}$ (Eq. \eqref{eq:MFA-LLB}) and $\alpha^{\rm{B}}_e$ from the coupling between the Gd and the Fe sublattices. 
After switching, $H_{\rm{Fe}}\approx H_{\rm{Gd}}$ and relativistic relaxation processes dominate the dynamics and drive magnetization to complete the switching. 
The question here is to what extent the non-equilibrium fields as given by Eq. \eqref{eq:effective-fields-MFA-Mentink} are accurate, and how are the relaxation parameters related to atomic damping parameters in ASD.

\begin{figure}[!tb]
\includegraphics[width=8.25cm]{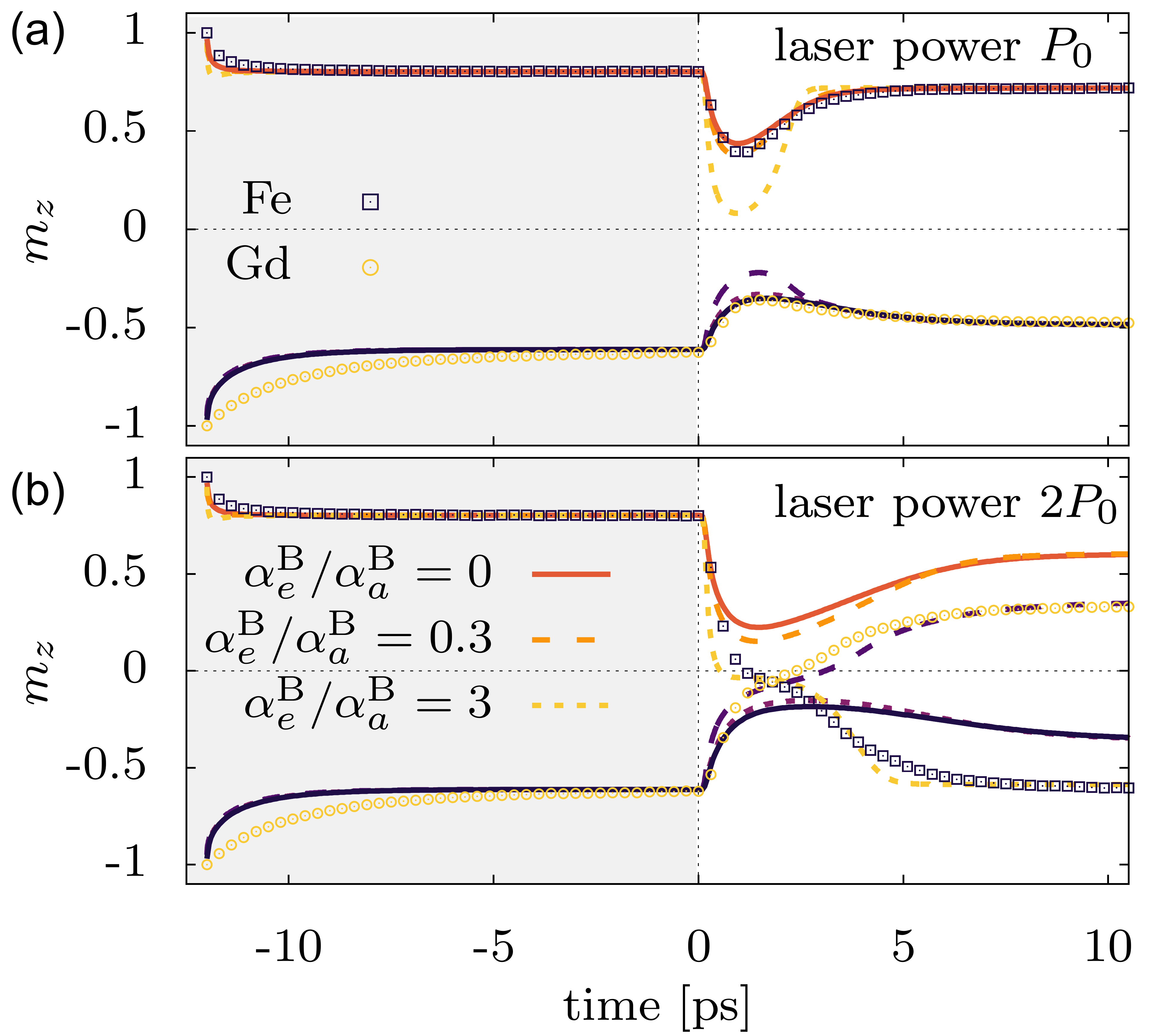}
\caption{Element-specific magnetization dynamics of GdFeCo calculated using atomistic spin dynamics (symbols) and macroscopic Baryakhtar-like equation (solid lines) for two laser pulse power values, (a) $P_0$ and (b) $2P_0$. Both electron and lattice temperature are kept constant, $T=300$ K, for  $t<0$. At $t=0$ a laser pulse is applied. In the Baryakhtar-like model the relativistic relaxation parameters $\alpha^{\rm{B}}_a$ have a value different to the Gilbert damping in ASD simulations, 
$ (\gamma/\mu_{\rm{Fe}})\alpha^{\rm{B}}_{\rm{Fe}}=0.005$ and $(\gamma/\mu_{\rm{Gd}})\alpha^{\rm{B}}_{\rm{Gd}}=0.01$.
The exchange relaxation parameter is varied, $\alpha^{\rm{B}}_e/\alpha^{\rm{B}}_{\rm{Fe}}=0,0.3$ and 3.
The relaxation to thermal state ($t<0$) is only well described for the Fe sublattice.
(a) For $P_0$, the laser induced dynamics is well described by  $\alpha^{\rm{B}}_e/\alpha^{\rm{B}}_{\rm{Fe}}=0.1$. 
(b) For $2P_0$ the demagnetization phase of both sublattices is relatively well described in comparison to ASD simulations. Switching is also possible, here one instance, for a value $\alpha^{\rm{B}}_e/\alpha^{\rm{B}}_{\rm{Fe}}=3$.}
\label{fig:ASD-Ba-comp}
\end{figure}

So far the connection between the relaxation parameters in the ASD and Baryakhtar-like model is unknown. 
In ASD simulations shown in Fig. \ref{fig:ASD-switching} we have used $\lambda_{\rm{Fe}}=\lambda_{\rm{Gd}}=0.01$ as atomistic relaxation parameter. One would expect that the relaxation parameters in the atomistic and macroscopic models are related as $\lambda_{a} \approx \alpha^{\rm{B}}_{a} (\gamma_a/\mu_{a})$.
In an attempt to find this correspondence, we directly compare results from ASD simulations and Baryakhtar-like models for different values of $\alpha^{\rm{B}}_a$ and $\alpha^{\rm{B}}_e$ in Eqs. \eqref{eq:Onsager-multisublattice-a} and \eqref{eq:Onsager-multisublattice-b}. 
 We numerically solve Eqs. \eqref{eq:Onsager-multisublattice-a},\eqref{eq:Onsager-multisublattice-b}, and \eqref{eq:effective-fields-MFA-Mentink} coupled to the 2TM with exactly the same parameters as for the ASD simulations. 
After exploring the results of the Baryakhtar model for a range of values for $\alpha^{\rm{B}}_a$ and $\alpha_e$, we find that for some values the agreement is good, as one observes in Fig. \ref{fig:ASD-Ba-comp}, however, it is not possible to find a good match for all scenarios.

 In order to illustrate this, we first focus on
 the dynamics induced by the laser pulse with power $P_0$ (Fig. \eqref{fig:ASD-Ba-comp}(a)).
 We find a good match for the laser induced magnetization dynamics ($t>0$ for  
 $ (\gamma/\mu_{\rm{Fe}})\alpha_{\rm{Fe}}=0.005$ and $(\gamma/\mu_{\rm{Gd}})\alpha_{\rm{Gd}}=0.01$, and for values of exchange relaxation of up to $\alpha^{\rm{B}}_e/\alpha^{\rm{B}}_{\rm{Fe}}=0.3$.
 For values $\alpha^{\rm{B}}_e/\alpha^{\rm{B}}_{\rm{Fe}}<0.3$, thermal relaxation ($t<0$) of the Fe  is also well described, however the relaxation of the Gd sublattice is significantly faster. For larger values of the exchange relaxation $\alpha^{\rm{B}}_e/\alpha^{\rm{B}}_{\rm{Fe}}=3$,  the dynamics of both sublatttices are substantially speed up and strongly disagree with ASD simulations. 
 

For larger laser pulse power $2P_0$ the magnetization switches using ASD simulations. We keep the same values for the relaxation parameters in Baryakhtar-like model as for $P_0$, and compare to the ASD simulations. 
 For small values of $\alpha^{\rm{B}}_{e}$ (Fig. \eqref{fig:ASD-Ba-comp}(b)), differently to the $P_0$ case (Fig. \eqref{fig:ASD-Ba-comp}(a)), the dynamics described by the Baryakhtar-like model is not only slower than those of ASD simulations but it hardly reproduces magnetization switching. 
 In order to reproduce switching,  we need to use larger values of the exchange relaxation parameter, $\alpha^{\rm{B}}_e/\alpha^{\rm{B}}_{\rm{Fe}}=3$.  
 These findings are in agreement with previous works using  Baryakhtar-like model where switching was reproduced for comparable values of $\alpha^{\rm{B}}_e$. However, as we have discussed before, for those values of  $\alpha^{\rm{B}}_e$, thermal relaxation dynamics ($t<0$) is much faster than in ASD simulations.
 This brings us to the question of how much understanding about switching can we gain by using this bare Baryakhtar-like model, are we missing something?

\subsection{The Landau-Lifshitz-Bloch model}
Since the Baryakhtar-like model is based on symmetry arguments, the macroscopic magnetization dynamics coming out from ASD simulations should also be described by that model with adequate expression for the relaxation parameters and non-equilibrium effective fields. 
 The magnetization dynamics coming out from ASD simulations is well described by the LLB equation of motion.
\begin{equation}
\frac{d m_a}{d t} = \Gamma_{\|,a} \left(m_a-m_{0,a} \right),
\label{eq:long-LLB}
\end{equation}
where 
\begin{equation}
\Gamma_{\|,a}=2  \lambda_a \frac{\gamma}{\mu_a} k_B T \frac{1}{\xi_a}\frac{L(\xi_a)}{L'(\xi_a)},
\label{eq:long-LLB-rate}
\end{equation}
 with $\xi_a=\beta \mu_a H_a^{\rm{MFA}}$, where $H_a^{\rm{MFA}}$ is given in Eq. \eqref{eq:MFA-LLB}, and  $m_{0,a} = L(\xi_a)$. 
The same equation applies to the second sublattice $b$.
Here, the relaxation rate $\Gamma_{\|,a}$ depends non-linearly on the non-equilibrium sublattice magnetization, $m_{a(b)}$, through the parameter $\xi_a$. 
We note that Eq. \eqref{eq:long-LLB} can be expanded around equilibrium for small perturbations of the magnetization. 
By doing so, the relaxation rates and effective fields are expressed in terms of equilibrium properties such as equilibrium magnetization and zero-field susceptibilities \cite{Atxitia2012}. 
In the present work, however, we use the version in Eq. \eqref{eq:long-LLB}. 
\begin{figure}[!tb]
\includegraphics[width=8.25cm]{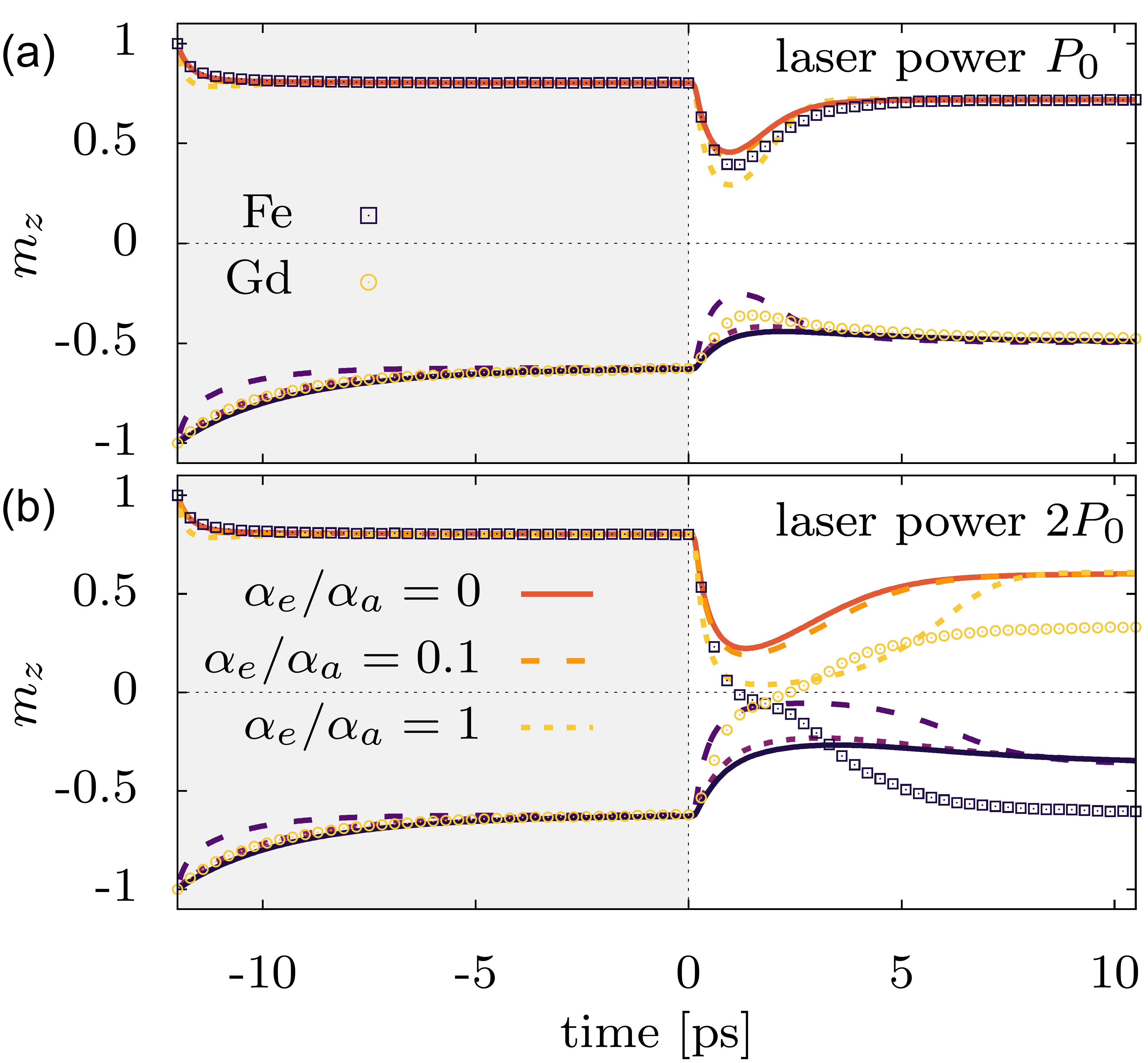}
\caption{Element-specific magnetization dynamics of GdFeCo calculated using atomistic spin dynamics (symbols) and macroscopic LLB equation (solid lines) for two laser pulse power values, (a) $P_0$ and (b) $2P_0$. For $t<0$, electron and lattice temperature are $T=300$K, and at $t=0$ a laser pulse is applied.
The exchange relaxation parameter is varied, $\alpha_e/\alpha_a=0,0.1$ and 1, where $\alpha_a=0.01$, and $a=$FeCo or Gd.  
The initial relaxation dynamics is well described by  $\alpha_e/\alpha_a=0$. 
(a) For laser power $P_0$, the element-specific dynamics is well-described for $\alpha_e/\alpha_a=0.1$. (a) For $\alpha_e/\alpha_a=1$, exchange relaxation dominates and the element-specific dynamics are similar. (b) For laser power $2P_0$, the switching dynamics is not described by the LLB model.}
\label{fig:ASD-LLB-comp}
\end{figure}
Direct comparison between ASD simulations and the LLB model of element-specific magnetization dynamics is possible and with relatively good agreement.  
Importantly, since the LLB model is derived directly from the ASD microscopic model, the damping parameters, $\lambda_{a(b)}$ in Eqs. \eqref{eq:long-LLB-rate} and \eqref{eq:sLLG} stand for the same physics, the rate of angular momentum dissipation of the atomic spins.  Differently to the Baraykhtar model where $\alpha_{a(b)}^{\rm{B}}$ is taken as a fitting parameter, within the LLB model the value of $\lambda_{a(b)}$ in Eq. \eqref{eq:long-LLB-rate} is the same as in the ASD simulations. 
A key difference between the Baryakhtar-like model and the LLB model is that in the latter an exchange relaxation term is missing. 
In order to find a meeting point between these phenomenological models, we rewrite Eq. \eqref{eq:long-LLB} in terms of a damping term multiplied by an effective field, 
\begin{equation}
\frac{d m_a}{d t} = \frac{2\lambda_a L(\xi_a)}{\xi_a} \frac{\gamma}{\mu_a} \frac{m_a-m_{0,a}}{\beta L'(\xi_a)} =  \gamma \alpha_{a}  H_a,
\label{eq:long-LLB-1}
\end{equation}
where 
\begin{equation}
\alpha_{a} = 2\lambda_a \frac{L(\xi_a)}{\xi_a}.
\label{eq:alpha-a}
\end{equation}
Differently to Baryakhtar-like model, in the LLB model, the relaxation parameter strongly depends on temperature and non-equilibrium sublattice magnetization through the thermal field, $\xi_a= \beta \mu_a H_a^{\rm{MFA}}$.
 At the same time, the non-equilibrium fields $\mu_a H_a$ within the LLB and Baryakhtar-like models differ. 
The effective field in the LLB model is defined as
\begin{equation}
\mu_a H_a = \frac{(m_a-m_{0,a})}{\beta L'(\xi_a)}.
\label{eq:muaHa}
\end{equation}
Equation \eqref{eq:muaHa} provides a microscopic description of the effective field driving the magnetization dynamics in ferrimagnets, based on the Heisenberg spin model (Eq. \eqref{eq:Ham}).
 Under the assumption of small perturbations around the equilibrium both, LLB and Baryakhtar-like effective fields, simplify to Landau-like expressions \cite{Mentink2017}. 
Equation \eqref{eq:long-LLB-1} describes with a very good degree of accuracy the relaxation of the angular momentum via dissipation to the heat-bath, which corresponds to the relativistic term in Eqs. \eqref{eq:Onsager-multisublattice-a}
and \eqref{eq:Onsager-multisublattice-b}. 
Previously, it has been found that ASD simulations compare well to Eq. \eqref{eq:long-LLB-1} for coupling parameters of around $\lambda_a\approx 0.1-1$ \cite{Atxitia2012,Nieves2014}. 
These values can be considered to correspond to the intermediate-to-high coupling regime. Direct comparison between ASD simulations and experiments of single pulse switching in GdFeCo has suggested values of $\lambda_{\rm{Fe}}\approx 0.06$ and $\lambda_{\rm{Gd}}\approx 0.01$ \cite{Jakobs2021}.   
In the context of the present work we find that Eq. \eqref{eq:long-LLB-1} describes relatively well the thermal relaxation dynamics in direct comparison to ASD simulations (Fig. \eqref{fig:ASD-LLB-comp}).

In order to account for the exchange relaxation in the LLB model, we follow the Baryakhtar-like model (\eqref{eq:Onsager-multisublattice-a} and \eqref{eq:Onsager-multisublattice-b}), and add an exchange relaxation term to Eq. \eqref{eq:long-LLB-1},
\begin{equation}
\frac{\mathrm{d} m_a}{\mathrm{d} t} =  \gamma \alpha_{a}  H_a + \gamma \frac{\alpha_{e}}{\mu_a} ( \mu_a H_a - \mu_b H_b )
\label{eq:long-LLB-1-ex}
\end{equation}
where $\alpha_{e}$ is a phenomenological exchange relaxation parameter to be determined by comparison to ASD dynamics. 
The inclusion of the exchange relaxation (second term in r.h.s) in the LLB improves the agreement to ASD simulations. 
With this addition, the LLB model describes well thermal relaxation for small values of the ratio $\alpha_{e}/\alpha_a$ as demonstrated in Fig. \ref{fig:ASD-LLB-comp}. For large values of $\alpha_e$ the LLB model is unable to describe thermal relaxation dynamics ($t<0$ in Fig. \ref{fig:ASD-LLB-comp}(a) and (b)).
For laser power $P_0$ (Fig. \ref{fig:ASD-LLB-comp}(a) ($t>0$)) the magnetization dynamics is slightly slower using the LLB model than those gained by ASD simulations for $\alpha_{e}/\alpha_a=0$. For $\alpha_{e}/\alpha_a=0.1$, the agreement is even better than without exchange relaxation.
The agreement vanishes when the exchange relaxation is increased to $\alpha_{e}/\alpha_a=1$. Critically, when the laser power is increased from $P_0$ to $2P_0$, for which ASD simulations show ultrafast switching, the LLB model only shows demagnetization-remagnetization of both sublattices. We find some agreement on the demagnetization time scales when a quite large exchange relaxation is used, $\alpha_{e}/\alpha_a=1$. These dynamics are similar to those observed using the Baryakhtar-like model for intermediate values of the exchange relaxation parameter (Fig. \eqref{fig:ASD-Ba-comp}). It has been demonstrated previously that  by including the transverse components of the equation of motion, switching is possible via a precessional path when a canting between the magnetization of each sublattice exists \cite{Atxitia2013}.  Here, we restrict to purely longitudinal switching within the LLB model.

\subsection{Unified phenomenological model}
\label{sec:secIV}
 
 \begin{figure}[!t]
\includegraphics[width=8.6cm]{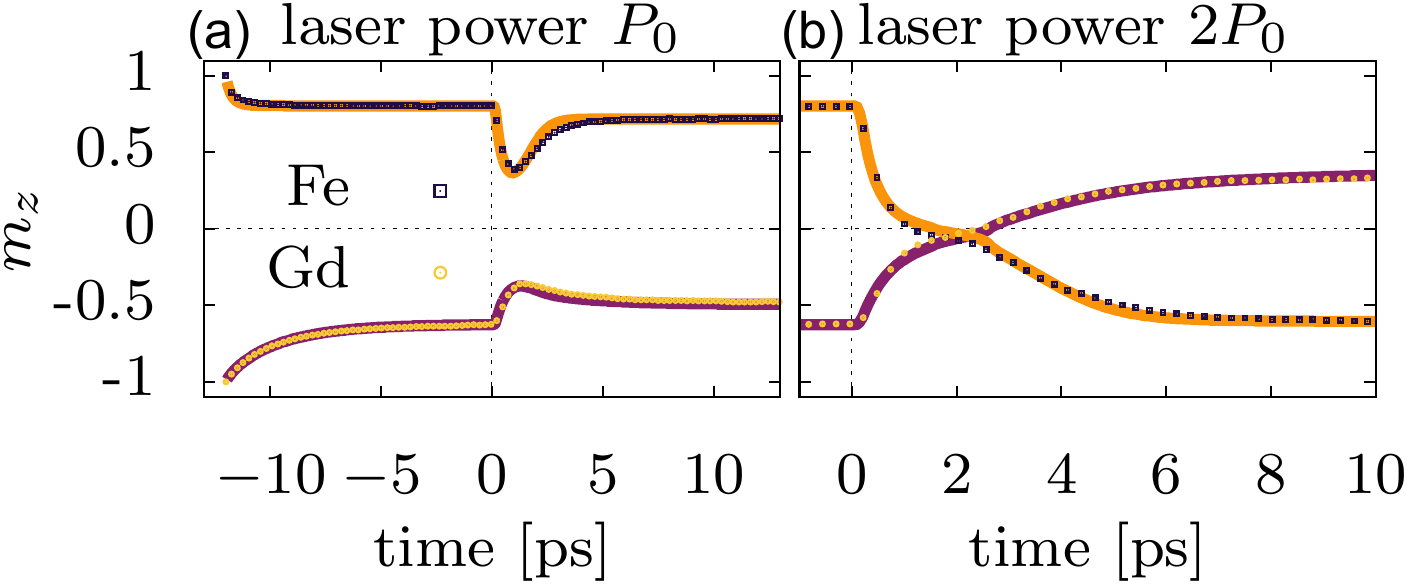}
\caption{Element-specific magnetization dynamics of GdFeCo calculated using atomistic spin dynamics (symbols) and the unified phenomenological model derived here, following Eq. \eqref{eq:long-LLB-1-ex} (solid lines) for two laser pulse power values, (a) $P_0$ and (b) $2P_0$. Both electron and lattice temperature are kept constant, $T=300$ K, for  $t<0$. At $t=0$ a laser pulse is applied, the same as in Figure \eqref{fig:ASD-switching}. 
GdFeCo alloys with $x_{\rm{Gd}}=25 \%$ are calculated.}
\label{fig:ASD-final-comparison}
\end{figure}

 So far we have constructed a phenomenological model based on the LLB and Baryakhtar-like models, the dynamics is given by Eq. \eqref{eq:long-LLB-1-ex}, the effective field by Eq. \eqref{eq:muaHa} and the relativistic relaxation parameter Eq. \eqref{eq:alpha-a}. We still need an expression for the exchange relaxation parameter.   
We construct this expression starting with  single species ferromagnets, where sublattices $a$ and $b$ represent the same spin lattice, hence exchange of angular momentum is non-local. Therefore,    $\mu_aH_a - \mu_bH_b = \mu_a H_{\rm{ex}} a_0^2 \Delta m_a$, with $a_0$ representing the lattice constant. Hence, the rate of non-local angular momentum transfer reads $\Gamma_{\rm{ex}}^{\rm{non-loc.}}=\alpha_{\rm{ex}} (\mu_a H_a - \mu_b H_b) = \alpha_a (A/M_a(T)) \Delta m_a$, where $A$ is the so-called micromagnetic exchange stiffness \cite{Atxitia2010}.  $M_a(T)=(\mu_a/\upsilon_a)m_a$ is the magnetization density at temperature $T$, where $\upsilon_a$ is the unit cell volume. Therefore, we find that $\alpha_{\rm{ex}}= \alpha_a/(z m_a)$. 
By considering that the exchange relaxation rate should conserve the symmetry under the exchange of lattice index, $\alpha_{ex}(M_1,M_2)=\alpha_{ex}(M_2,M_1)$, 
 we find that 
\begin{equation}
\alpha_{\rm{ex}} = \frac{1}{2}\left(\frac{\alpha_{a}}{z_{ab} m_a} +  \frac{\alpha_{b}}{z_{ba} m_b} \right).
\label{eq:alpha-ex}
\end{equation}
This expression is the extension of the non-local exchange relaxation in ferromagnets to local exchange relaxation in ferrimagnets.
This explicit expression for the exchange relaxation parameter in Eq. \eqref{eq:alpha-ex} completes our unified model, which bridges the atomistic spin dynamics model and the Baryakhtar and LLB macroscopic models. 

We find that the agreement between our unified phenomenological model and ASD simulations is excellent, see Fig. \eqref{fig:ASD-final-comparison}(a) and (b). 
Figure \ref{fig:ASD-final-comparison}(a) shows that for $t<0$, the sublattice magnetization relaxation towards thermal equilibrium value is described with a high level of accuracy by our model. For $t>0$ and a relatively low laser power $P_0$, the agreement is also excellent for the demagnetization and remagnetization dynamics. 
Figure \ref{fig:ASD-final-comparison}(b) shows the comparison between the unified model and ASD simulations of the switching dynamics.
We conclude that Eq. \eqref{eq:long-LLB-1-ex} for the sublattice magnetization dynamics together with the Eq. \eqref{eq:muaHa} for the effective field and Eqs. \eqref{eq:alpha-a} and \eqref{eq:alpha-ex} for the relaxation parameters, unify the Barayakhtar and the LLB phenomenological models for single-pulse all-optical switching in ferrimagnets.





\section{Discussion and conclusion}
\label{sec:secV}

The macroscopic model presented in this work solves some open questions in the field of ultrafast magnetization dynamics in ferrimagnets. For example, it answers the question of the range of applicability and the validity of the parameters of the Barayakhtar and LLB phenomenological models. 
In the one hand, within our model, the relativistic relaxation parameters ($\alpha_a$) are element-specific and strongly depend on both the temperature and the non-equilibrium sublattice magnetization. 
The temperature and magnetization dependence of the relativistic relaxation parameters are well described by the LLB model.
In the other hand,  
the exchange relaxation parameter ($\alpha_{\rm{ex}}$) is cast in terms of the element specific relativistic relaxation parameters and sublattice magnetization. 
We have demonstrated that in order to reproduce the ASD simulations results, the relaxation parameters in the Barayakhtar model have to be both temperature and magnetization dependent. 
The explicit expression of the exchange relaxation parameter is the main result of the present work since it allows us to unify the Barayakhtar and LLB models.
While for the Barayakhtar model $\alpha_{e}$ is unconnected to $\alpha_a$, within our proposed model they are proportional to each other, $\alpha_e \sim \alpha_a/m_a$. This relation is the key to bridge both ASD simulations and Barayakhtar and LLB models together. 
Additionally, we have also demonstrated the validity of the non-equilibrium effective fields given in Eq. \eqref{eq:muaHa} as derived in the LLB model instead of the Barayakhtar model.


Single-pulse switching in ferrimagnets has been described before by the Baryakhtar model. A necessary condition for switching is that the system transits from the relativistic relaxation regime to the so-called exchange-dominated relaxation regime. 
Although details of switching in such a regime have been already discussed in detail \cite{Mentink2012,Mentink2017}, so far it has remained unknown how this transition could be described theoretically.
Our model resolves this question. When the system is at equilibrium or weakly excited, the exchange-relaxation parameter fulfills, 
$\alpha_{e}\ll \alpha_{a}$. For strong excitation, such that the magnetic order of one sublattice reduces significantly, close to zero $m_a \rightarrow 0$, the exchange relaxation will dominate the dynamics since $\alpha_{e} \sim \alpha_a /m_a \gg \alpha_{a}$.
From our model, one can derive universal criteria for switching in ferrimagnets, including GdFeCo and Mn$_2$Ru$_x$Ga \cite{Jakobs2022arxiv}.

The provided understanding is paramount for further research on material engineering, for example, to find alternative material classes showing all-optical switching. Notably, our model predicts that the exchange relaxation term is enhanced as the number of neighbours reduces. This dependence suggests that magnetic systems of lower dimension, e.g. 2D magnets \cite{review2D}, could show a faster, more efficient switching than bulk materials. 
Further, the extension of our model to the micromagnetic level will allow to optimize switching conditions. 
The use of micromagnetic computational solvers permits for a realistic description of ultrafast AOS processes in ferrimagnetic alloys, such as helicity-independent and helicity-dependent AOS, where multidomain states and thermal gradients play an important role in the process \cite{RaposoPRB2022}.




To summarize, in the present work we have presented 
a unified model for single-pulse all-optical switching in ferrimagnets.
Our model merges and improves previous semi-phenomenological models, the Landau-Lifshitz-Bloch model and Barayakhtar-like models. 
To verify the accuracy of the proposed model, we directly compare the laser induced magnetization dynamics to atomistic spin dynamics computer simulations. Differently to previous models, 
our model has the advantage that it can be directly compared to ASD simulations.  
Further, we have established the connection between ASD and macroscopic equations of motion.
Importantly, we provide here the stepping stone for the construction of a micromagnetic model valid for ferrimagnets including exchange relaxation between sublattices.
This is paramount for a robust construction of a multiscale scheme of the switching process in which not only local magnetization dynamics is described but also magnetic domain nucleation and motion under strong non-equilibrium.
Multiscale-based micromagnetic models will allow for the description of realistic sample sizes and describe recent spintronics phenomena using laser pulses, e.g. magnetic skyrmion creation/deletion with fs laser pulses, or domain-wall motion under dynamics thermal gradients.

\begin{acknowledgments}
 The authors acknowledge support from the Deutsche Forschungsgemeinschaft through SFB/TRR 227  "Ultrafast Spin Dynamics", Project A08.
\end{acknowledgments}

\bibliography{library}

\end{document}